\documentclass[11pt]{article}
\usepackage[dvips]{graphics}
\title{Covariant canonical quantization of fields
and Bohmian mechanics}
\author{Hrvoje Nikoli\'c \\
Theoretical Physics Division, Rudjer Bo\v{s}kovi\'{c} Institute, \\
P.O.B. 180, HR-10002 Zagreb, Croatia \\
{\normalsize e-mail: hrvoje@thphys.irb.hr} \\
\makebox[1in]{} \\
}
\date{\today}
\begin{document}
\maketitle
\begin{abstract}
We propose a manifestly covariant canonical 
method of field quantization
based on the classical De Donder-Weyl covariant 
canonical formulation of field theory.
Owing to covariance, the space and time arguments of fields are treated 
on an equal footing. To achieve both covariance and 
consistency with standard noncovariant canonical
quantization of fields in Minkowski spacetime, it is necessary 
to adopt a covariant Bohmian formulation of
quantum field theory. A preferred foliation 
of spacetime emerges dynamically owing to a purely quantum effect. 
The application to a simple time-reparametrization invariant system 
and quantum gravity is discussed and compared with the conventional
noncovariant Wheeler-DeWitt approach.    
\end{abstract}

%\pacs{04.20.Fy, 04.60.Ds, 04.60.Gw, 04.60.-m}
%Canonical formalism, Lagrangians, and variational principles
%Canonical quantization
%Covariant and sum-over-histories quantization
%Quantum gravity

%\noindent
%PACS: 04.20.Fy; 04.60.Ds; 04.60.Gw; 04.60.-m \\
%Keywords: Quantum field theory; Canonical quantization; Covariance; 
%Bohmian mechanics; Quantum gravity

\section{Introduction} 

One of the main open problems of modern theoretical physics is 
how to combine the principle of quantum mechanics with  
the principle of general relativity. One aspect of this problem 
- the existence of nonrenormalizable infinities in 
perturbative quantum gravity - seems to be solved by two major approaches
to quantum gravity, namely, by string theory \cite{gsw} and 
by loop quantum gravity \cite{rovlect,thiem,rovbook}. Nevertheless, 
neither of these two approaches fully incorporates the principle 
of general relativity. String theory is a perturbative approach 
depending on the choice of the background metric, while 
loop quantum gravity is a canonical approach that does not treat 
time on an equal footing with space. There are attempts to solve 
these problems by introducing a background independent M-theory 
for strings \cite{banks} or a spacetime spinfoam formalism 
for loops \cite{rovbook}, but these attempts are not yet fully 
successful. 

An alternative, in a sense more conservative approach 
is to try to modify the usual canonical quantization rules
for fields (where time plays a special role) by introducing  
{\em covariant} canonical quantization rules. In fact, 
several classical covariant canonical formalisms are 
already known, such as the covariant phase space formalism 
\cite{crnwitt,crn} and various versions of the multimomenta 
formalism (see, e.g., \cite{kas,got,rov} and references 
therein). Unfortunately, a satisfying method of quantization 
based on some of these classical formalisms is still not 
known. Perhaps, the most towards covariant canonical 
quantization of fields has been done in \cite{kan}, 
where a method of quantization of the classical De Donder-Weyl 
covariant canonical formalism has been proposed. 
(For the application to quantum gravity, see \cite{kan2}.)
However, this proposal suffers from two problems. First, the 
quantum formalism introduces a new fundamental dimensional 
constant, the physical meaning of which is not completely clear. 
Second, the theory is able to describe the usual time-space asymmetric 
wave-functional states $\Psi([\phi({\bf x})],t)$, but only
if $\Psi([\phi({\bf x})],t)$ is a product state of the form 
$\prod_{{\bf x}}\Psi(\phi({\bf x}),{\bf x},t)$. The aim of 
this paper is to propose a different method of quantization 
based on the classical De Donder-Weyl formalism, such that  
the problems of the approach of \cite{kan} are avoided.

Before starting with the De Donder-Weyl formalism, 
let us give a few additional qualitative remarks on possibilities 
for covariant quantization. The classical covariant phase space 
method \cite{crnwitt,crn} rests on the existence of space of 
classical solutions to the equations of motion. However, in the 
conventional formulation of quantum field theory, it is not clear 
what might play the role of an analog of classical solutions. 
On the other hand, in the deterministic Bohmian formulation 
of quantum field 
theory \cite{bohm2,bohmrep,holrep,holbook}, such an analog 
exists. This suggests that the Bohmian formulation might be a 
basis for a covariant formulation of quantum field theory.
In fact, among various equivalent formulations of 
nonrelativistic quantum mechanics \cite{ajp}, 
the Bohmian formulation is conceptually the most similar to classical 
physics, which again suggests that well-understood classical 
covariant theories might be covariantly quantized most easily 
by using the Bohmian formulation. Finally, since space and time should play 
equal roles in a covariant quantum field theory, in general, 
one might expect wave functionals of the form $\Psi([\phi(x)],x)$ 
instead of the usual time-space asymmetric
wave functionals $\Psi([\phi({\bf x})],t)$. This implies that 
time-dependent fields $\phi(x)=\phi({\bf x},t)$ should play a role 
in a covariant approach,
which might be related 
to the Bohmian formulation that assigns a deterministic time evolution 
to the field $\phi({\bf x},t)$. 

Of course, the arguments above 
%(see also \cite{nikqg})
have only a heuristic value and are
not sufficiently convincing by themselves.
However, in the present paper, the heuristic arguments above are 
turned into much stronger and more convincing arguments.
In contrast to the usual approaches to Bohmian mechanics, 
the Bohmian formulation is {\em not postulated} for interpretational 
purposes, but {\em derived} from the purely technical 
requirements - covariance and consistency with standard 
quantum field theory. In this way, a covariant version of Bohmian 
mechanics emerges automatically, as a part of the formalism without which the
theory cannot be formulated consistently. This, together with the results of 
\cite{nikol1,nikol2} related to relativistic first quantization, 
suggests that it is Bohmian mechanics that might be the missing 
bridge between quantum mechanics and relativity. In addition, 
we also note that the Bohmian interpretation might play an important 
role for quantum cosmology \cite{bar,col,pn,mar,pn2} 
and noncommutative theories \cite{barb}. 

The paper is organized as follows. Sec.~\ref{prel} contains two 
subsections, one presenting a
short review of the classical covariant canonical 
De Donder-Weyl formalism, while the other presenting a short review 
of the Bohmian formulation of the conventional canonical 
field quantization. For simplicity, the results are presented 
for a real scalar field in flat spacetime.
The central section, Sec.~\ref{Quant}, 
combines the results presented in Sec.~\ref{prel} to formulate 
a covariant canonical quantum formalism consistent with 
the conventional canonical quantization of fields in flat 
spacetime. Sec.~\ref{Gener} contains generalizations that 
include a larger number of fields and curved spacetime.
In Sec.~\ref{Reparinv}, the formalism 
is applied to a simple toy model 
with time-reparametrization invariance, as well as to quantum 
gravity. The discussion of our results is presented in the 
final section, Sec.~\ref{Disc}.

In the paper, we use the units such that the velocity of light is 
$c=1$, while the signature of the metric is $(+,-,-,-)$.

\section{Preliminaries}
\label{prel}

\subsection{Classical De Donder-Weyl formalism}

In this subsection we briefly review the classical 
covariant canonical De Donder-Weyl formalism. (For more details, 
we refer the reader to \cite{kas,rieth} and references therein.)
For simplicity, we present the formalism for one real scalar field
in Minkowski spacetime, while the generalizations are discussed in
Sec.~\ref{Gener}. 

Let $\phi(x)$ be a real scalar field described by the 
action 
\begin{equation}\label{action}
{\cal A}=\int d^4x {\cal L}, 
\end{equation}
where
\begin{equation}
{\cal L}=\frac{1}{2} (\partial^{\mu}\phi)(\partial_{\mu}\phi) 
- V(\phi) .
\end{equation}
The corresponding covariant canonical momentum is given by the 4-vector
\begin{equation}
\pi^{\mu}=\frac{\partial {\cal L}}{\partial (\partial_{\mu}\phi)}
=\partial^{\mu}\phi .
\end{equation}
The covariant canonical equations of motion are
\begin{equation}\label{caneqm}
\partial_{\mu}\phi=\frac{\partial {\cal H}}{\partial \pi^{\mu}},
\;\;\;\;\;    
\partial_{\mu}\pi^{\mu}=-\frac{\partial {\cal H}}{\partial\phi},
\end{equation}
where the scalar De Donder-Weyl Hamiltonian (not related to the energy
density!) is given by the Legendre transform 
\begin{eqnarray}\label{Hcovar}
{\cal H}(\pi^{\alpha}, \phi) 
 & = & \pi^{\mu}\partial_{\mu}\phi-{\cal L} \nonumber \\
 & = & \frac{1}{2} \pi^{\mu}\pi_{\mu} + V.
\end{eqnarray}
Eqs.~(\ref{caneqm})
are equivalent to the standard Euler-Lagrange equations of motion.
By introducing the local vector $S^{\mu}(\phi(x),x)$, the dynamics 
can also be described by the covariant De Donder-Weyl 
Hamilton-Jacobi equation
\begin{equation}\label{DWHJ}
{\cal H}\left( \frac{\partial S^{\alpha}}{\partial\phi}, \phi \right)
+\partial_{\mu}S^{\mu}=0 ,
\end{equation}
together with the equation of motion
\begin{equation}\label{eqmHJ}
\partial^{\mu}\phi=\pi^{\mu}=\frac{\partial S^{\mu}}{\partial\phi}.
\end{equation}
Note that in (\ref{DWHJ}), 
$\partial_{\mu}$ is the {\em partial} derivative 
acting only on the second argument of $S^{\mu}(\phi(x),x)$. 
The corresponding total derivative is given by
\begin{equation}\label{totder}
d_{\mu}=\partial_{\mu}+(\partial_{\mu}\phi)\frac{\partial}{\partial \phi}.
\end{equation}   
Note also that (\ref{DWHJ}) is a single equation for four quantities 
$S^{\mu}$. Consequently, there is a lot of freedom in finding 
solutions to (\ref{DWHJ}). Nevertheless, the theory is equivalent 
to other formulations of classical field theory.

Now, following \cite{kan}, we consider the relation between 
the covariant Hamilton-Jacobi equation (\ref{DWHJ}) and
the conventional Hamilton-Jacobi equation. The latter can be 
derived from the former in the following way. 
Using (\ref{Hcovar}), (\ref{DWHJ}) takes the explicit form
\begin{equation}\label{DWHJ2}
\frac{1}{2} \frac{\partial S_{\mu}}{\partial\phi}
\frac{\partial S^{\mu}}{\partial\phi}  + V + \partial_{\mu}S^{\mu}=0.
\end{equation}
Using the equation of motion (\ref{eqmHJ}), we write the 
first term in (\ref{DWHJ2}) as
\begin{equation}\label{pom1} 
\frac{1}{2} \frac{\partial S_{\mu}}{\partial\phi}
\frac{\partial S^{\mu}}{\partial\phi} =
\frac{1}{2} \frac{\partial S^0}{\partial\phi}
\frac{\partial S^0}{\partial\phi} +
\frac{1}{2} (\partial_i \phi)(\partial^i \phi) ,
\end{equation}
where $i=1,2,3$ are the space indices. Similarly, using 
(\ref{eqmHJ}) and (\ref{totder}), we write the
last term in (\ref{DWHJ2}) as
\begin{equation}\label{pom2}
\partial_{\mu}S^{\mu}=\partial_0S^0 +d_iS^i
-(\partial_i \phi)(\partial^i \phi).
\end{equation}
Now introduce the quantity
\begin{equation}\label{Scal}
{\cal S}=\int d^3x\, S^0,
\end{equation}
so that
\begin{equation}\label{pom3}
\frac{\partial S^0(\phi(x),x)}{\partial\phi(x)}=
\frac{\delta {\cal S}([\phi({\bf x},t)],t)}{\delta\phi({\bf x};t)},
\end{equation}
where
\begin{equation}\label{funcder}
\frac{\delta}{\delta\phi({\bf x};t)} \equiv 
\left.
\frac{\delta}{\delta\phi({\bf x})} 
\right|_{\phi({\bf x})=\phi({\bf x},t)} 
\end{equation}
is the space functional derivative.
Thus, putting (\ref{pom1}), (\ref{pom2}), and (\ref{pom3})  
into (\ref{DWHJ2}) and integrating the resulting equation over $d^3x$, 
we obtain 
\begin{equation}\label{HJs}
\int d^3x \left[ \frac{1}{2}
\left( \frac{\delta {\cal S}}{\delta\phi({\bf x};t)} 
\right)^2
+\frac{1}{2}(\nabla \phi)^2 +V(\phi) \right] +\partial_t {\cal S} =0 ,
\end{equation}
which is the standard noncovariant Hamilton-Jacobi equation
(written for the time-dependent field $\phi({\bf x},t)$).
The time evolution of the field $\phi({\bf x},t)$ is given by 
\begin{equation}\label{eqmcan}
\partial_t \phi({\bf x},t)=\frac{\delta {\cal S}}{\delta\phi({\bf x};t)},
\end{equation}
which is a consequence of the time component of (\ref{eqmHJ}).
Note that in deriving (\ref{HJs}) from (\ref{DWHJ2}), it was necessary 
to use the space part of the equations of motion (\ref{eqmHJ}). 
Whereas this fact does not play an important role in classical physics, 
it has far reaching 
consequences in the quantum case studied in Sec.~\ref{Quant}. 

\subsection{Bohmian formulation of quantum field theory}

Quantum field theory can be formulated in the functional 
Schr\"odinger picture as
\begin{equation}\label{sch}
\hat{H}\Psi=i\hbar\partial_t\Psi,
\end{equation}
where, for the real scalar field $\phi$,  
\begin{equation}
\hat{H}=
\int d^3x \left[ -\frac{\hbar^2}{2}
\left( \frac{\delta}{\delta\phi({\bf x})}
\right)^2 
+\frac{1}{2}(\nabla \phi)^2 +V(\phi) \right].
\end{equation}
By writing
\begin{equation}\label{wf}
\Psi([\phi({\bf x})],t)={\cal R}([\phi({\bf x})],t)
e^{i{\cal S}([\phi({\bf x})],t)/\hbar},
\end{equation}
where ${\cal R}$ and ${\cal S}$ are real functionals,
one finds that the complex equation (\ref{sch}) is 
equivalent to a set of two real equations
\begin{equation}\label{QHJs}
\int d^3x \left[ \frac{1}{2}
\left( \frac{\delta {\cal S}}{\delta\phi({\bf x})}
\right)^2
+\frac{1}{2}(\nabla \phi)^2 +V(\phi) +Q
\right]  +\partial_t {\cal S} =0 ,
\end{equation} 
\begin{equation}\label{conssch2}
\int d^3x \left[ \frac{\delta{\cal R}}{\delta\phi({\bf x})}
\frac{\delta{\cal S}}{\delta\phi({\bf x})} +J \right] +\partial_t {\cal R}
=0,
\end{equation}
where
\begin{equation}\label{Qpot}
Q=-\frac{\hbar^2}{2{\cal R}}  
\frac{\delta^2 {\cal R}}{\delta\phi^2({\bf x})} ,
\end{equation}
\begin{equation}\label{J}
J=\frac{{\cal R}}{2}
\frac{\delta^2 {\cal S}}{\delta\phi^2({\bf x})}  .
\end{equation}
Eq.~(\ref{conssch2}) is also equivalent to
\begin{equation}\label{conssch}
\partial_t {\cal R}^2 + \int d^3x \frac{\delta}{\delta\phi({\bf x})}
\left( {\cal R}^2 \frac{\delta {\cal S}}{\delta\phi({\bf x})} \right) =0 .
\end{equation}
Eq.~(\ref{conssch}) represents the unitarity of the theory, 
because it provides that the norm
\begin{equation}
\int [d\phi({\bf x})]\, \Psi^*\Psi=\int [d\phi({\bf x})]\, {\cal R}^2
\end{equation}
does not depend on time. The quantity 
${\cal R}^2 ([\phi({\bf x})],t)$ represents the probability density 
for fields to have the configuration $\phi({\bf x})$ at time $t$. 
Instead of starting with (\ref{sch}), one can equivalently 
take Eqs.~(\ref{QHJs}) 
and (\ref{conssch2}) as the starting point for quantization of fields.
In addition, since $\Psi$ must be a single-valued quantity, 
in the approach based on (\ref{QHJs})                                    
and (\ref{conssch2}) one must require that the quantity 
$\exp{i{\cal S}/\hbar}$ should be single valued.

Eqs.~(\ref{QHJs}) and (\ref{conssch}) also suggest an interesting 
interpretation, known as the Bohmian interpretation of 
quantum field theory \cite{bohm2,bohmrep,holrep,holbook}. 
The interpretation consists in the assumption that quantum fields 
have a deterministic time evolution given by the classical 
equation (\ref{eqmcan}). Remarkably, the statistical predictions 
of this deterministic interpretation are equivalent to those 
of the conventional interpretation. In the
deterministic  interpretation, 
all quantum uncertainties are a consequence of the ignorance 
of the actual initial field configuration $\phi({\bf x},t_0)$. 
The main reason for the consistency of this interpretation is the 
fact that (\ref{conssch}) with (\ref{eqmcan}) represents the 
continuity equation, which provides that the statistical distribution
$\rho([\phi({\bf x})],t)$ 
of field configurations $\phi({\bf x})$ is given by 
the quantum distribution $\rho={\cal R}^2$ at {\em any} time $t$, 
provided that $\rho$ is given by ${\cal R}^2$ at some initial 
time $t_0$. The initial distribution is arbitrary in principle, but a 
quantum H-theorem \cite{val} explains why the quantum distribution 
is the most probable.

Comparing (\ref{QHJs}) with (\ref{HJs}), we see that the quantum field 
satisfies an equation similar to the classical one, except 
for an additional quantum force resulting from the {\em nonlocal} 
quantum potential $Q$. (The nonlocality implies that the 
Bohmian hidden variable theory, with $\phi({\bf x},t)$ being 
the hidden variable, is not in contradiction with the 
Bell theorem that asserts that {\em local} hidden variables 
cannot be consistent with quantum mechanics.)  
The quantum equation of motion turns out to be
\begin{equation}\label{emb}
\partial^{\mu}\partial_{\mu}\phi+\frac{\partial V(\phi)}{\partial\phi} 
+\frac{\delta {\cal Q}}{\delta\phi({\bf x};t)} =0 ,
\end{equation}
where ${\cal Q}=\int d^3x\, Q$. The last term represents the deviation 
from the classical equation of motion. 

The Bohmian interpretation may seem appealing to some, while 
unattractive to the others. An appealing feature is an explanation 
of the notorious ``wave function collapse". An unattractive 
feature is the fact that the Bohmian interpretation 
is not covariant and requires 
the existence of a preferred Lorentz frame not determined by the theory, 
despite the fact that 
the statistical predictions obtained by averaging over hidden 
variables are Lorentz invariant \cite{holrep}. However, 
at this level, nothing forces us to adopt this interpretation, 
just as nothing prevents us from adopting it. 
Adopting or rejecting this interpretation is more like a matter of 
taste. As we shall see 
in the next section, in the manifestly covariant formulation 
of quantum field theory based on the De Donder-Weyl formalism,
the situation is quite different. There, 
one {\em must} adopt a covariant version of the Bohmian 
equations of motion, because otherwise one cannot retain both 
covariance and consistency with the standard canonical 
quantization in Minkowski spacetime.        

\section{Covariant canonical quantization}
\label{Quant}

Our basic idea for covariant canonical quantization is to find 
a quantum substitute for the classical covariant 
De Donder-Weyl Hamilton-Jacobi equation (\ref{DWHJ2}). 
For this purpose, we first formulate the classical De Donder-Weyl 
theory in a slightly different, more general way. Let $A([\phi],x)$ 
be a functional of $\phi(x)$ and a function of $x$.
We define the derivative 
\begin{equation}\label{derd}
\frac{dA([\phi],x)}{d\phi(x)}\equiv \int d^4x'
\frac{\delta A([\phi],x')}{\delta\phi(x)},
\end{equation}
where $\delta/\delta\phi(x)$ is the spacetime functional derivative
(not the space functional derivative in (\ref{funcder})). 
In particular, if $A([\phi],x)$ is a local functional, i.e., if 
$A([\phi],x)=A(\phi(x),x)$, then 
\begin{equation}\label{func1}
\frac{dA(\phi(x),x)}{d\phi(x)}= 
\int d^4x'\frac{\delta A(\phi(x'),x')}{\delta\phi(x)}
=\frac{\partial A(\phi(x),x)}{\partial\phi(x)}.
\end{equation}
Thus we see that the derivative $d/d\phi$ 
is a generalization of the ordinary 
partial derivative $\partial/\partial\phi$, such that its action 
on nonlocal functionals is also well defined. An example
of a particular interest 
is a functional nonlocal in space but local in time, 
so that
\begin{equation}
\frac{\delta A([\phi],x')}{\delta\phi(x)}=
\frac{\delta A([\phi],x')}{\delta\phi({\bf x};x^0)} \delta(x'^0-x^0) .
\end{equation}
In this case, one can write
\begin{equation}\label{func2}
\frac{dA([\phi],x)}{d\phi(x)}=\frac{\delta}{\delta\phi({\bf x};x^0)}
\int d^3x' A([\phi],{\bf x}',x^0).
\end{equation}
Being equiped with these mathematical tools, we can write
(\ref{DWHJ2}) as
\begin{equation}\label{DWHJ2nl}
\frac{1}{2} \frac{d S_{\mu}}{d\phi}
\frac{d S^{\mu}}{d\phi}  + V + \partial_{\mu}S^{\mu}=0,
\end{equation} 
which is the form appropriate for the quantum modification.
Similarly, the classical equations of motion (\ref{eqmHJ}) can be written
as
\begin{equation}\label{eqmHJnl}
\partial^{\mu}\phi=\frac{d S^{\mu}}{d\phi}.
\end{equation} 

Now we are ready to propose a method of quantization that combines 
the classical covariant canonical De Donder-Weyl formalism with 
the standard time-space asymmetric canonical quantization of fields. 
Our starting point is the relation between the noncovariant classical 
Hamilton-Jacobi equation (\ref{HJs}) and its quantum analog 
(\ref{QHJs}). Suppressing the time dependence of the field in (\ref{HJs}),
we see that 
they differ only in the existence of the $Q$-term 
in the quantum case. This suggests us to postulate the following quantum 
analog of the classical covariant equation (\ref{DWHJ2nl}): 
\begin{equation}\label{QDWHJ2}    
\frac{1}{2} \frac{d S_{\mu}}{d\phi}
\frac{d S^{\mu}}{d\phi}  + V+Q + \partial_{\mu}S^{\mu}=0.
\end{equation}
Here $S^{\mu}=S^{\mu}([\phi],x)$ is a functional (not merely a function) 
of $\phi(x)$. This means that $S^{\mu}$ at $x$ may depend on 
the field $\phi(x')$ at {\em all} points $x'$. Such 
spacetime nonlocalities absent in classical physics are expected  
in quantum physics. Indeed, space nonlocalities appear 
in the conventional time-space asymmetric quantum 
field theory. Here, for the sake 
of covariance, we also allow the time nonlocalities (see also \cite{nikol2}).  
Thus Eq.~(\ref{QDWHJ2}) is manifestly 
covariant, provided that $Q$ given by (\ref{Qpot}) can be written 
in a covariant form. The quantum equation (\ref{QDWHJ2}) 
must be consistent with the conventional quantum equation 
(\ref{QHJs}). Indeed, by using a similar procedure to that 
used to show that (\ref{DWHJ2}) implies (\ref{HJs}),
one can show that (\ref{QDWHJ2}) implies 
(\ref{QHJs}), provided that some additional conditions are fulfilled.
First, $S^0$ must be local in time, so that (\ref{func2}) for 
$A=S^0$ can be used (compare with (\ref{pom3})).
Second, $S^i$ must be completely local, so that 
$dS^i/d\phi=\partial S^i/\partial\phi$, which implies
\begin{equation}\label{totderQ}
d_iS^i=\partial_iS^i+(\partial_i\phi)\frac{d S^i}{d \phi}
\end{equation}
(compare with (\ref{totder})).
However, just as in the classical case, 
in this procedure it is {\em necessary}
to use the space part of the equations of motion (\ref{eqmHJ}). 
Therefore, these classical equations of motion must be valid even in the 
quantum case. Since we want a covariant 
theory in which space and time play equal roles, the validity 
of the space part of the equations of motion (\ref{eqmHJ}) implies 
that their time part should also be valid. Consequently, in the covariant 
quantum theory based on the De Donder-Weyl formalism, one must require
the validity of (\ref{eqmHJnl}).
This requirement is nothing but a covariant version of the Bohmian equation 
of motion, written for an arbitrarily nonlocal $S^{\mu}$.
(Note that, in order to achieve the consistency of the
De Donder-Weyl quantization with the conventional quantization,
the space part of the
classical equations of motion have also been used in the approach
of \cite{kan}. However, in that paper, the physical consequences of this
fact have not been recognized.)

The next step is to find a covariant substitute for 
Eq.~(\ref{conssch2}). For this purpose, we introduce a vector 
$R^{\mu}([\phi],x)$.  
The vector field $R^{\mu}$ can be viewed as
generating a preferred foliation of spacetime, such 
that, in this foliation,
the vector $R^{\mu}$ is normal to the leafs of that foliation.
This allows us to introduce the quantity
\begin{equation}\label{Rcal}
{\cal R}([\phi],\Sigma)=\int_{\Sigma} d\Sigma_{\mu}R^{\mu},
\end{equation}
where $\Sigma$ is a leaf (a 3-dimensional hypersurface) generated 
by $R^{\mu}$. Similarly, a covariant version of (\ref{Scal}) 
reads
\begin{equation}\label{Scalc}
{\cal S}([\phi],\Sigma)=\int_{\Sigma} d\Sigma_{\mu}S^{\mu},
\end{equation}
where $\Sigma$ is generated by $R^{\mu}$ again. 
Consequenly, the covariant version
of (\ref{wf}) reads 
\begin{equation}\label{wfc}
\Psi([\phi],\Sigma)={\cal R}([\phi],\Sigma)
e^{i{\cal S}([\phi],\Sigma)/\hbar}.
\end{equation}
For $R^{\mu}$ we postulate the equation
\begin{equation}\label{Rmu}
\frac{dR^{\mu}}{d\phi}\frac{dS_{\mu}}{d\phi} +J 
+\partial_{\mu}R^{\mu}=0 .
\end{equation}
In this way, a preferred foliation emerges
dynamically, as a foliation generated by the
solution $R^{\mu}$ of the equations (\ref{Rmu}) and (\ref{QDWHJ2}).
Note that $R^{\mu}$ does not play any role in classical physics,
so the existence of a preferred foliation is a purely quantum effect.
Now the relation between (\ref{Rmu}) and (\ref{conssch2}) is obtained 
by assuming that nature has chosen 
a solution of the form $R^{\mu}=(R^0,0,0,0)$, where $R^0$ is local in time. 
In this case, 
by integrating (\ref{Rmu}) over $d^3x$
and assuming again that $S^0$ is local in time,
one obtains (\ref{conssch2}).
Thus we see that (\ref{Rmu}) is a covariant substitute for (\ref{conssch2}).

It remains to write covariant versions of (\ref{Qpot}) and (\ref{J}). 
They are simply
\begin{equation}\label{Qpotc}
Q=-\frac{\hbar^2}{2{\cal R}}
\frac{\delta^2 {\cal R}}{\delta_{\Sigma}\phi^2(x)} ,
\end{equation}
\begin{equation}\label{Jc}
J=\frac{{\cal R}}{2}
\frac{\delta^2 {\cal S}}{\delta_{\Sigma}\phi^2(x)} ,
\end{equation}
where $\delta/\delta_{\Sigma}\phi(x)$ is a version of
(\ref{funcder}) in which $\Sigma$ is generated by $R^{\mu}$.
Here $\Sigma$ depends on $x$ (the point $x$ is an element 
of $\Sigma$) and $\Sigma$ is  
kept fixed in the variation $\delta_{\Sigma}\phi(x)$. 
Thus, (\ref{Rmu}) with (\ref{Jc})
and (\ref{QDWHJ2}) with (\ref{Qpotc}) represent 
a covariant substitute for the functional Schr\"odinger equation 
(\ref{sch}) equivalent to (\ref{conssch2}) with (\ref{J})
and (\ref{QHJs}) with (\ref{Qpot}).

The covariant Bohmian equations (\ref{eqmHJnl}) imply a covariant version 
of (\ref{emb})
\begin{equation}\label{embc}
\partial^{\mu}\partial_{\mu}\phi+\frac{\partial V}{\partial\phi}
+\frac{d Q}{d\phi} =0 .
\end{equation}
Since the last term can also be written as
$\delta(\int d^4x Q)/\delta\phi(x)$, the equation of motion (\ref{embc})
can be obtained by varying the quantum action 
\begin{equation}\label{QA}
{\cal A}_Q=\int d^4x {\cal L}_Q=\int d^4x ({\cal L}-Q).
\end{equation}

To summarize, our covariant canonical quantization of fields is given by 
Eqs.~(\ref{QDWHJ2}), (\ref{Rmu}), (\ref{Qpotc}), (\ref{Jc}), 
and (\ref{eqmHJnl}). 
The conventional functional Schr\"odinger equation corresponds to a 
special class of solutions of (\ref{QDWHJ2}), (\ref{Rmu}), (\ref{Qpotc}) and 
(\ref{Jc}), for which $R^i=0$, $S^i$ are local,  
while $R^0$ and $S^0$ are local in time.

\section{Generalizations}
\label{Gener}

In this section we generalize the results obtained so far to include 
the cases of a larger number of fields, as well as the case of 
curved spacetime (that may be either a background spacetime or 
a dynamical spacetime). 

Let $\phi(x)=\{ \phi_a(x) \}$ be a collection of fields. 
We study a classical action (\ref{action}),
with ${\cal L}$ taking the form
\begin{eqnarray}\label{Lgen}
{\cal L} &=& \frac{1}{2} G^{ab}(\phi,x)g^{\mu\nu}(x)
(\partial_{\mu}\phi_a)(\partial_{\nu}\phi_b) \\ \nonumber
& & +F^{a\mu}(\phi,x)\partial_{\mu}\phi_a -V(\phi,x).
\end{eqnarray}
In particular, $G^{ab}$, $F^{a\mu}$, and $V$ are proportional 
to $|g|^{1/2}$, where $g$ is the determinant of the metric tensor 
$g_{\mu\nu}$. Thus the factor $|g|^{1/2}$ is included in the 
definition of ${\cal L}$, which makes the application of
canonical methods easier. (The price we pay is that the 
general covariance is slightly less manifest.)
We also introduce the quantity $G_{ab}$ defined by 
\begin{equation}
G_{ab}G^{bc}=\delta_a^c.
\end{equation}
Thus $G_{ab}$ is the matrix inverse to $G^{ab}$.
(For the case in which the inverse does not exist,
see below.) 
This allows us 
to consistently raise and lower the indices 
$a,b$ with $G^{ab}$ and 
$G_{ab}$, respectively. 
Since $G^{ab}\propto |g|^{1/2}$, we see that if 
$\partial_{\mu}\phi_a$ is a tensor, 
then $\partial^{\mu}\phi^a$ is a tensor density.
% of weight $-1$  
%(or simply - tensor density) \cite{wein}.
 
Now the canonical momenta are
\begin{equation}
\pi^{a\mu}=\frac{\partial {\cal L}}{\partial(\partial_{\mu}\phi_a)}
=\partial^{\mu}\phi^a +F^{a\mu},
\end{equation}  
while the De Donder-Weyl Hamiltonian is
\begin{eqnarray}
{\cal H} &=& \pi^{a\mu}\partial_{\mu}\phi_a -{\cal L} \\ \nonumber
 &=& \frac{1}{2} (\partial^{\mu}\phi^a)(\partial_{\mu}\phi_a) +V
     \\ \nonumber
 &=& \frac{1}{2} \pi^{a\mu}\pi_{a\mu} -\pi^{a\mu}F_{a\mu}
     + \frac{1}{2} F^{a\mu}F_{a\mu} +V.
\end{eqnarray}
The corresponding covariant canonical equations of motion are
\begin{equation}
\partial_{\mu}\phi_a = \frac{\partial {\cal H}}{\partial\pi^{a\mu}}
=\pi_{a\mu}-F_{a\mu},
\end{equation}
\begin{equation}
\partial^{\mu}\pi_{a\mu}=-G_{ab}\frac{\partial {\cal H}}{\partial\phi_b}
\equiv -\partial_a {\cal H}.
\end{equation}
Here $\partial_a= G_{ab}\partial^b \neq \partial^bG_{ab}$, because
$G_{ab}$ depends on $\phi$.
The covariant Hamilton-Jacobi equations are
\begin{equation}
\pi^{a\mu}=\frac{\partial S^{\mu}}{\partial\phi_a}\equiv\partial^aS^{\mu},
\end{equation}
\begin{equation}\label{DWHJgen}
\frac{1}{2}(\partial^aS^{\mu})(\partial_aS_{\mu})-F_{a\mu}\partial^aS^{\mu}
+\frac{1}{2}F^{a\mu}F_{a\mu}+V +\partial_{\mu}S^{\mu}=0.
\end{equation}
The total derivative is
\begin{equation}
d_{\mu}=\partial_{\mu}+(\partial_{\mu}\phi_a)\partial^a .
\end{equation} 

It is instructive to show explicitly that (\ref{DWHJgen}) is general 
covariant. It is not difficult to see that each term in 
(\ref{DWHJgen}) is proportional to $|g|^{1/2}$, i.e., that the 
left-hand side is a scalar density. In particular, $S^{\mu}$ is a 
vector density, so we write $S^{\mu}=|g|^{1/2}\tilde{S}^{\mu}$, where
$\tilde{S}^{\mu}$ is a vector. To obtain a scalar equation, we multiply 
the whole equation with $|g|^{-1/2}$. The last term becomes 
$|g|^{-1/2}\partial_{\mu}(|g|^{1/2}\tilde{S}^{\mu})$, which is nothing 
but the covariant derivative $\nabla_{\mu}\tilde{S}^{\mu}$. In a similar 
way, one can show that all other equations of this section 
are also general covariant.

The general covariant generalisation
of the derivative (\ref{derd}) depends on the tensor nature of $A$.   
In our case $A$ is a vector density $A^{\mu}$, so (\ref{derd})
naturally generalizes to
\begin{equation}\label{derdcov}
\frac{dA^{\mu}([\phi],x)}{d\phi(x)} \equiv 
\frac{e^{\mu}_{\bar{\alpha}}(x)}{|g(x)|^{1/2}}
\int d^4x'\, e_{\nu}^{\bar{\alpha}}(x')
\frac{\delta A^{\nu}([\phi],x')}{\delta\phi(x)} .
\end{equation}
Here $e^{\mu}_{\bar{\alpha}}$ is the tetrad satisfying
$e^{\mu}_{\bar{\alpha}}e^{\bar{\alpha}\nu}=g^{\mu\nu}$, 
where $\bar{\alpha}$ is an index in the internal $SO(1,3)$ 
group.

Now the quantization is straightforward. In (\ref{DWHJgen}), one 
replaces the derivative $\partial^a$ with the derivative 
$d^a=d/d\phi_a$ and adds the $Q$-term. 
The quantum potential is
\begin{equation}\label{Qpotcg}
Q=-\frac{\hbar^2}{2{\cal R}}
\frac{\delta}{\delta_{\Sigma}\phi_a}G_{ab}\frac{\delta}{\delta_{\Sigma}\phi_b}
{\cal R} .
\end{equation}
Eq.~(\ref{Rmu}) generalizes to
\begin{equation}\label{Rmug}
(d^aR^{\mu})(d_aS_{\mu})-F_{a\mu}d^aR^{\mu} +J
+\partial_{\mu}R^{\mu}=0 ,
\end{equation}
where
\begin{equation}\label{Jcg}
J=\frac{{\cal R}}{2}
\frac{\delta}{\delta_{\Sigma}\phi_a} \left(
G_{ab}\frac{\delta{\cal S}}{\delta_{\Sigma}\phi_b} -F_{a\mu}r^{\mu} \right),
\end{equation}
and $r^{\mu}=R^{\mu}/(R^{\lambda}R_{\lambda})^{1/2}$.
The particular orderings in (\ref{Qpotcg}) 
and (\ref{Jcg}) are chosen so that they lead to a
Schr\"odinger equation with a hermitian Hamilton operator.
Eqs.~(\ref{Rcal}) and (\ref{Scalc}) now take 
the manifestly covariant form
\begin{equation}\label{Rcalg}
{\cal R}([\phi],\Sigma)=\int_{\Sigma} d\Sigma_{\mu}\tilde{R}^{\mu},
\end{equation}
\begin{equation}\label{Scalg}
{\cal S}([\phi],\Sigma)=\int_{\Sigma} d\Sigma_{\mu}\tilde{S}^{\mu},
\end{equation}   
where $\tilde{S}^{\mu}$ and $\tilde{R}^{\mu}=R^{\mu}/|g|^{1/2}$ are vectors.
The wave functional is again given by (\ref{wfc}) and 
the Bohmian equations of motion 
\begin{equation}
\partial_{\mu}\phi_a = d_aS_{\mu}-F_{a\mu}
\end{equation}
are equivalent to the equations obtained by varying the 
quantum action (\ref{QA}).

The formalism can be further generalized to the case in which
$G^{ab}g^{\mu\nu}$ in (\ref{Lgen}) is replaced with a more 
general quantity of the form $G^{ab\mu\nu}$. This, of course, 
does not present any problem for the classical formalism.
The quantum equation (\ref{Rmug}) generalises to
\begin{equation}\label{Rmug2}
G_{ab\mu\nu}\frac{dR^{\mu}}{d\phi_a}\frac{dS^{\nu}}{d\phi_b}
-G_{ab\mu\nu}F^{b\nu}\frac{dR^{\mu}}{d\phi_a} +J
+\partial_{\mu}R^{\mu}=0 ,
\end{equation}   
where $G_{ab\mu\nu}$ is the inverse of $G^{ab\mu\nu}$, in the sense that
\begin{equation}
G^{ab'\mu\nu'}G_{b'b\nu'\nu}=\delta^a_b\delta^{\mu}_{\nu}.
\end{equation}
The generalizion of (\ref{Qpotcg}) and (\ref{Jcg}) 
consists in the replacement of  
$G_{ab}$ in (\ref{Qpotcg}) and (\ref{Jcg})
with $G^{(r)}_{ab}$, where
\begin{equation} 
G^{(r)}_{ab}=G_{ab\mu\nu}r^{\mu}r^{\nu} .
\end{equation}
By considering the case $R^{\mu}=(R^0,0,0,0)$, 
it is easy to see that this generalization leads to the usual 
Schr\"odinger equation.

In some cases, such as gauge theories, the invers of 
$G^{ab\mu\nu}$ does not exist. In such cases,
one can define $G^{ab\mu\nu}$
and its inverse $G_{ab\mu\nu}$ by 
introducing a gauge fixing term or by using 
other tricks \cite{kas,hor,kan3}.

Our formalism leads to particularly 
interesting consequences when applied to 
quantum gravity and other theories with reparametrization invariance.  
We study this in more detail in the next section.

\section{Reparametrization-invariant theories}
\label{Reparinv}

\subsection{A toy model}

Consider a system with two degrees of freedom $\phi_1(t)$ and 
$\phi_2(t)$. Let the action be given by
\begin{equation}\label{actiontoy}
A=\int dt \, L =\int dt \, \phi_2 \left[ \frac{\dot{\phi_1}^2}{2(\phi_2)^2}
-\tilde{V}(\phi_1) \right], 
\end{equation}
where the dot denotes the time derivative $d_t$. The action is 
invariant with respect to time reparametrizations $t\rightarrow t'(t)$, 
$\phi_2\rightarrow\phi'_2=(dt/dt')\phi_2$. As is well known, 
such theories lead to a Hamiltonian constraint and thus serve as  
toy models instructive for an easier understanding of some of the peculiar
properties of classical and quantum gravity \cite{kuc,padm}. 
Here we study this model by using 
the Hamilton-Jacobi formalism. 
Since there is no space (but only time) in this model, the 
De Donder-Weyl canonical formalism is identical to the conventional
classical canonical formalism.

The canonical momenta are
\begin{equation}
\pi^1=\frac{\partial L}{\partial \dot{\phi_1}}=\frac{\dot{\phi_1}}{\phi_2},
\;\;\;\;
\pi^2=\frac{\partial L}{\partial \dot{\phi_2}}=0.
\end{equation}
Since $\pi^2=0$, the Hamiltonian is
\begin{equation}
H=\pi^1\dot{\phi_1}-L=\phi_2\left( \frac{(\pi^1)^2}{2}+\tilde{V} \right) .    
\end{equation}
The Hamilton-Jacobi formalism is given by the equations of motion
\begin{equation}\label{HJemtoy}
\pi^1=\frac{\partial S}{\partial \phi_1}=\frac{\dot{\phi_1}}{\phi_2}, 
\;\;\;\;
\pi^2=\frac{\partial S}{\partial \phi_2}=0,
\end{equation}
together with the Hamilton-Jacobi equation 
\begin{equation}\label{HJtoy}
\phi_2\left[ \frac{1}{2}\left(\frac{\partial S}{\partial \phi_1}\right)^2
+\tilde{V} \right] +\partial_t S =0.
\end{equation}
From the second equation in (\ref{HJemtoy}), we see 
that $S$ does not depend on $\phi_2$. Consequently, by applying
the derivative $\partial/\partial\phi_2$ to (\ref{HJtoy}), 
one obtains
\begin{equation}\label{Hcons} 
\frac{1}{2}\left(\frac{\partial S}{\partial \phi_1}\right)^2
+\tilde{V}=0,
\end{equation}
which is nothing but the Hamiltonian constraint $H=0$. It is equivalent 
to the equation of motion that one obtains by varying (\ref{actiontoy}) 
with respect 
to $\phi_2$. Comparing (\ref{Hcons}) with (\ref{HJtoy}), one 
also finds
\begin{equation}\label{Stoy}
\partial_t S =0.
\end{equation}
Applying the derivative $\partial/\partial\phi_1$ to (\ref{Hcons}), 
one obtains
\begin{equation}\label{pomtoy}
\frac{\partial S}{\partial \phi_1} \frac{\partial }{\partial \phi_1}
\frac{\partial S}{\partial \phi_1} 
+\frac{\partial\tilde{V}}{\partial \phi_1}=0.
\end{equation}
Using (\ref{HJemtoy}) and the fact that
\begin{equation}\label{totdertoy}
\dot{\phi_1}\frac{\partial }{\partial \phi_1} \frac{\partial S}{\partial
\phi_1} =d_t \frac{\partial S}{\partial \phi_1},
\end{equation}
(\ref{pomtoy}) implies 
\begin{equation}
d_t \left( \frac{\dot{\phi_1}}{\phi_2} \right) 
+\frac{\partial\tilde{V}}{\partial\phi_1} =0.
\end{equation}
This is nothing but the equation of motion obtained by varying 
$\phi_1$ in (\ref{actiontoy}).

Now consider the quantization. Following the general method 
developed is Sec.~\ref{Gener}, we introduce the quantum potential 
\begin{equation}\label{pottoy}
Q=-\frac{\hbar^2}{2R}\phi_2\frac{d^2R}{d\phi_1^2},
\end{equation}
and replace (\ref{HJtoy}) with 
\begin{equation}\label{HJtoyQ}
\phi_2 \left[ \frac{1}{2}\left(\frac{d S}{d \phi_1}\right)^2
+\tilde{V} \right] +Q +\partial_t S =0.
\end{equation}
Eqs.~(\ref{HJemtoy}) are valid with the derivatives
$\partial/\partial\phi_a$ replaced with $d/d\phi_a$. 
The conservation equation reads
\begin{equation}\label{constoy}
\partial_tR^2 +\frac{d}{d\phi_1}\left( R^2\phi_2
\frac{dS}{d\phi_1} \right)=0.
\end{equation}
Applying the derivative $d/d\phi_2$ to (\ref{HJtoyQ}), 
we obtain
\begin{equation}\label{HconsQ}
\frac{1}{2}\left(\frac{d S}{d \phi_1}\right)^2
+\tilde{V}+\tilde{Q}+\phi_2\frac{d\tilde{Q}}{d\phi_2}=0,
\end{equation}
where $Q=\phi_2\tilde{Q}$.
Combining (\ref{HJtoyQ}) and (\ref{HconsQ}), we obtain
\begin{equation}\label{StoyQ}
\partial_t S -(\phi_2)^2\frac{d\tilde{Q}}{d\phi_2}=0.
\end{equation}
Eqs.~(\ref{HconsQ}) and (\ref{StoyQ}) are the quantum 
analogs of (\ref{Hcons}) and (\ref{Stoy}), respectively.

Now compare the results above with the conventional 
method of quantization of the Hamiltonian constraint, 
based on the Wheeler-DeWitt equation $\hat{H}\Psi=0$. 
By writing $\Psi=R\exp(iS/\hbar)$, one obtains
\begin{equation}\label{HconsWDW}
\frac{1}{2}\left(\frac{d S}{d \phi_1}\right)^2
+\tilde{V}+\tilde{Q} =0,
\end{equation}
\begin{equation}\label{StoyWDW}
\partial_t S=0,
\end{equation}
and
\begin{equation}\label{consWDW}
\frac{d}{d\phi_1}\left( R^2
\frac{dS}{d\phi_1} \right)=0,
\end{equation}
instead of (\ref{HconsQ}),
(\ref{StoyQ}), and (\ref{constoy}),
respectively.
In particular, we see that $S$ and $R$ are time independent,
which corresponds to the 
well-known problem of time in quantum gravity 
\cite{padm,kuc2,ish}, consisting in the fact that 
the wave function(al) $\Psi$ does not depend on time. In our 
approach, $\partial_t S \neq 0$ and $\partial_t R \neq 0$
in the general quantum case, 
so there is no problem of time. However, our quantization 
contains the Wheeler-DeWitt quantization as a special 
case. If $R$ is a time-independent solution, then  
(\ref{constoy}) reduces to (\ref{consWDW}). Consequently, 
$R$ may be a $\phi_2$-independent solution of (\ref{constoy}). 
If $R$ does not depend on $\phi_2$, then (\ref{pottoy}) implies 
$d\tilde{Q}/d\phi_2=0$.
Consequently, (\ref{HconsQ}) and (\ref{StoyQ}) 
reduce to (\ref{HconsWDW}) and (\ref{StoyWDW}), respectively.

It is also instructive to show explicitly that our method of quantization 
leads to the Bohmian equations of motion that can be derived from 
the Lagrangian $L_Q=L-Q$, or equivalently, from the Hamiltonian
\begin{equation}
H_Q=H+Q .
\end{equation}
We have
\begin{eqnarray}\label{pomtoy1}
& \dot{\phi}_1=\displaystyle\frac{dH_Q}{d\pi^1}=\phi_2\pi^1, & \nonumber \\
& \dot{\pi}^1=-\displaystyle\frac{dH_Q}{d\phi_1}=-\phi_2 \left(
\frac{\partial\tilde{V}}{\partial\phi_1}
+\frac{d\tilde{Q}}{d\phi_1} \right),
\end{eqnarray}
\begin{eqnarray}\label{pomtoy2}
& \dot{\phi}_2=\displaystyle\frac{dH_Q}{d\pi^2}=0, & \nonumber \\
& 0=\dot{\pi}^2=-\displaystyle\frac{dH_Q}{d\phi_2}=
\frac{(\pi^1)^2}{2} +\tilde{V} +\tilde{Q}
+\phi_2\frac{d\tilde{Q}}{d\phi_2} . &
\end{eqnarray}
The second equation in (\ref{pomtoy2}) is equivalent to (\ref{HconsQ}).
Therefore, the only nontrivial Hamilton equation that remains 
to be proved in the quantum Hamilton-Jacobi framework 
is the second equation in (\ref{pomtoy1}).
Applying the derivative $d/d\phi_1$ to (\ref{HJtoyQ}) and using
(\ref{HJemtoy}) and 
\begin{equation}\label{totdertoyQ}
\partial_t \frac{d S}{d \phi_1}+
\dot{\phi_1}\frac{d }{d \phi_1} \frac{d S}{d
\phi_1} =d_t \frac{d S}{d \phi_1}
\end{equation} 
(compare with (\ref{totdertoy})), one obtains
\begin{equation}
d_t \pi^1+\phi_2 \left(
\frac{\partial\tilde{V}}{\partial\phi_1}
+\frac{d\tilde{Q}}{d\phi_1} \right)=0,
\end{equation}
which is the second equation in (\ref{pomtoy1}).

Finally, note that (\ref{HJtoyQ}) and (\ref{constoy}) can also be derived 
from the Schr\"odinger equation 
\begin{equation}\label{S83}
\hat{H}\Psi=i\hbar\partial_t\Psi .
\end{equation}
At first sight, this seems to be inconsistent with the classical 
Hamiltonian constraint $H=0$. However, in the classical
limit $\hbar\rightarrow 0$, (\ref{S83}) leads to (\ref{HJtoy}), 
which, as we have seen, {\em does} lead to the Hamiltonian constraint 
(\ref{Hcons}). Thus, we have a remarkable result that (\ref{S83})
{\em is} consistent with $H=0$, provided that the Bohmian equation
corresponding to the second equation in (\ref{HJemtoy}) is valid.

\subsection{Quantum gravity}

In this subsection, we sketch the main points
relevant for the application to quantum gravity. 

The classical gravitational action is 
\begin{equation}
{\cal A}=\int d^4x |g|^{1/2} R,
\end{equation}
where $R$ is the scalar curvature.
To write the Lagrangian in a form appropriate for a canonical treatment, 
we write
\begin{equation}
|g|^{1/2}R=\frac{1}{2}G^{\alpha\beta\mu\gamma\delta\nu}
(\partial_{\mu}g_{\alpha\beta})(\partial_{\nu}g_{\gamma\delta})
+{\rm total \;\; derivative},
\end{equation}
and ignore the total-derivative term.
The quantity $G^{\alpha\beta\mu\gamma\delta\nu}$ and its 
inverse $G_{\alpha\beta\mu\gamma\delta\nu}$ depend
on $g_{\alpha\beta}$ but not on the derivatives of $g_{\alpha\beta}$
\cite{hor,grant,padmrep}.
%Here
%\begin{eqnarray}\label{Gup}
%G^{\alpha\beta\mu\gamma\delta\nu} &=& \frac{|g|^{1/2}}{2}
%[g^{\mu\nu}(g^{\alpha\beta}g^{\gamma\delta}-g^{\alpha\gamma}g^{\beta\delta})
%\nonumber \\
%& & +2g^{\mu\delta}(g^{\alpha\gamma}g^{\beta\nu}-g^{\nu\gamma}g^{\beta\alpha})].
%\end{eqnarray} 
The fields $\phi_a$ are the components of the metric $g_{\alpha\beta}$.
The essential property of our covariant canonical quantization 
is that all 10 components $g_{\alpha\beta}$ are quantized, in contrast 
to the conventional noncovariant canonical quantization where only 
the space components $g_{ij}$ are quantized. Following the general 
method developed in Sec.~\ref{Gener}, one finds the
following quantum equations:
\begin{equation}
\frac{1}{2}G_{\alpha\beta\mu\gamma\delta\nu}
\frac{dS^{\mu}}{dg_{\alpha\beta}} \frac{dS^{\nu}}{dg_{\gamma\delta}}
+Q +\partial_{\mu}S^{\mu}=0,
\end{equation}
\begin{equation}
Q=-\frac{\hbar^2}{2{\cal R}}
\frac{\delta}{\delta_{\Sigma}g_{\alpha\beta}}
G^{(r)}_{\alpha\beta\gamma\delta}
\frac{\delta}{\delta_{\Sigma}g_{\gamma\delta}}
{\cal R} ,
\end{equation}
\begin{equation}
G_{\alpha\beta\mu\gamma\delta\nu}
\frac{dR^{\mu}}{dg_{\alpha\beta}} \frac{dS^{\nu}}{dg_{\gamma\delta}}
+J +\partial_{\mu}R^{\mu}=0,
\end{equation} 
\begin{equation}
J=\frac{{\cal R}}{2}
\frac{\delta}{\delta_{\Sigma}g_{\alpha\beta}}
G^{(r)}_{\alpha\beta\gamma\delta}
\frac{\delta}{\delta_{\Sigma}g_{\gamma\delta}}
{\cal S} ,
\end{equation}
where
\begin{equation}
G^{(r)}_{\alpha\beta\gamma\delta}=
G_{\alpha\beta\mu\gamma\delta\nu}r^{\mu}r^{\nu}.
\end{equation}
%\begin{equation}
%G^{\alpha\beta\mu\gamma'\delta'\nu'}G_{\gamma'\delta'\nu'\gamma\delta\nu}
%=\delta^{\alpha}_{\gamma}\delta^{\beta}_{\delta}\delta^{\mu}_{\nu}.
%\end{equation}

The Bohmian equations of motion 
\begin{equation}
\partial_{\mu}g_{\alpha\beta}=G_{\alpha\beta\mu\gamma\delta\nu}
\frac{dS^{\mu}}{dg_{\gamma\delta}}
\end{equation}
are equivalent to the equations of motion obtained
by varying the quantum action
\begin{equation}
{\cal A}_Q=\int d^4x (|g|^{1/2} R-Q).
\end{equation}
This leads to the equation of motion
\begin{equation}\label{EinstQ}
R^{\mu\nu}-\frac{g^{\mu\nu}}{2}R+|g|^{-1/2}\frac{dQ}{dg_{\mu\nu}}=0.
\end{equation}
The potential $Q$ is a scalar density, so we can write
$Q=|g|^{1/2}\tilde{Q}$, where $\tilde{Q}$ is a scalar. 
Consequently, (\ref{EinstQ}) can be written as
\begin{equation}\label{EinstQ2}
R^{\mu\nu}+\frac{d\tilde{Q}}{dg_{\mu\nu}}
-\frac{g^{\mu\nu}}{2}(R-\tilde{Q})=0.
\end{equation}
Another suggestive form is
\begin{equation}\label{EinstQ3}
\frac{g^{\mu\nu}}{2}R-R^{\mu\nu}=8\pi G_N T^{\mu\nu},
\end{equation}
where
\begin{equation}\label{Tqg}
T^{\mu\nu}=\frac{1}{16\pi G_N} \left( 2\frac{d\tilde{Q}}{dg_{\mu\nu}}
+g^{\mu\nu}\tilde{Q} \right).
\end{equation}

Note that (\ref{EinstQ3}) and (\ref{Tqg}) imply that the Bohmian 
equations of motion are fully covariant. By contrast, if 
the quantization of gravity is based on the 
conventional canonical Wheeler-DeWitt equation
that does {\em not} treat space and time on an equal footing, 
then the Bohmian interpretation leads to an equation similar 
to (\ref{EinstQ3}), but 
with a noncovariant energy-momentum tensor of the form \cite{shoj}
\begin{equation} 
T^{ij}\propto \frac{d\tilde{Q}}{dg_{ij}},
\;\;\;\;
T^{0\mu}\propto \tilde{Q}g^{0\mu}.
\end{equation} 
We also note that, similarly to the toy model studied in the 
preceding subsection, the conventional Wheeler-DeWitt quantization
correponds to a special case in which 
$R^i=0$, $S^i$ are local, while $S^0$ and $R^0$ are functionals 
which are local in time and do not depend on $g_{0\mu}$ and $x^0$. 

The functionals $S^{\mu}$ and $R^{\mu}$ (that determine 
also $\Psi$) describing the quantum state are functionals of the 
metric $g_{\alpha\beta}$. However, the theory is covariant, so 
one expects that physical results should not depend on the choice 
of coordinates, but only on the 4-geometry. In other words,
the theory is expected to be invariant with respect to
active 4-diffeomorhisms.  
At the moment, we do not know how to incorporate this property 
explicitly. However, if this 4-diffeomorhism invariance is 
explicitly realized, then one expects that ultraviolet divergences 
should be absent \cite{thiem}. Perhaps this could be realized 
explicitly by introducing a time-space symmetric version of 
loop quantum gravity based 
on the spacetime covariant Ashtekar variables
\cite{espos}.  

\section{Discussion}
\label{Disc}

The theory we have proposed in this paper offers a solution
to several fundamental problems, but also rises some new problems.

First, the theory offers a manifestly covariant method of 
field quantization, based on the classical De Donder-Weyl formalism.
The method treats space and time on an equal footing. Unlike 
the conventional canonical quantization, it is not formulated 
in terms of a single complex Schr\"odinger-like
equation, but in terms of two coupled real equations.
(In such a formulation, operators and commutation 
relations do not play any fundamental role.)
Nevertheless, if the solution satisfies certain additional conditions,
then the solution satisfies the conventional 
functional Schr\"odinger equation. 
These conditions are $R^i=0$, 
locality of $S^i$, and locality in time of $R^0$ and $S^0$. These 
conditions are clearly not time-space symmetric. 
Since the predictions of the Schr\"odinger  
equation in Minkowski spacetime are in agreement with experiments,
one would like to explain why nature chooses solutions that, 
at least approximately, satisfy these conditions.
Our theory does not explain that. However, in our theory, 
the observed time-space asymmetry of quantum field theory 
becomes a problem analogous to the observed time-space asymmetry
in cosmology (at large scales, the universe is homogeneous and 
isotropic in space but not in time), or to the observed time-space
asymmetry of thermodynamics (the entropy increases with time but not 
with space). 
Nothing prevents solutions that obey the observed
quantum, cosmological, or thermodynamical 
time-space asymmetric
rules, but a compelling explanation of these rules is missing.
This suggests that the observed quantum time-space asymmetry might be 
of the cosmological origin. Another possibility is that
our covariant quantum theory should be reformulated
such that the direction of $R^{\mu}$ fully determines the directions
in which the relevant quantities are local/nonlocal,
but this would make the theory less
elegant and perhaps would suppress interesting nonlocal
effects that might exist in nature.
(For example, such nonlocal effects
might play a role for solving the black-hole information paradox.)

Second, the theory offers a solution to the problem of time in 
quantum gravity.
In general, in our approach to quantum gravity, 
the functionals $S^{\mu}$ and $R^{\mu}$ depend 
on both space and time. The corresponding wave functional $\Psi$ 
depends on the hypersurface. Nevertheless, in the classical limit, 
the Hamiltonian constraint is always valid. In addition, if the 
quantum state satisfies certain additional conditions, then 
it satisfies the conventional Wheeler-DeWitt equation (with 
a fixed ordering of operators!). 

Third, the theory offers a covariant version of Bohmian 
mechanics, with time and space treated on an 
equal footing. The covariance of Bohmian mechanics is a direct 
consequence of the manifest covariance of the quantization procedure.
Note that a frequent argument against the previous versions 
of Bohmian mechanics is their dependence on the choice of 
the time coordinate, even when the predictions of the conventional 
interpretation of quantum field theory do not depend on this choice.
In these previous versions of Bohmian mechanics, one has to choose 
a preferred foliation of spacetime in a more or less ad hoc way 
(for a recent interesting attempt, see \cite{hort}). In our approach, 
the preferred foliation is generated dynamically by  
$R^{\mu}$. This quantity does not play any role in classical physics, 
so the preferred foliation is a purely quantum effect.
However, the Bohmian equations of motion themselves 
have a manifestly covariant form and do not depend 
explicitly on $R^{\mu}$.

Fourth, our theory offers a new reason why one should adopt 
the Bohmian interpretation. Of course, our covariant method of quantization 
by itself, just as any other method of quantization, 
does not automatically imply the Bohmian interpretation. However, 
the need for the Bohmian interpretation emerges from the requirement 
that our covariant method should be consistent with the 
conventional noncovariant method. Those who, for some 
personal reasons, do not like the Bohmian deterministic interpretation, 
may take this as a problem of our quantization method.  
In our view, this result (see also the heuristic arguments 
presented in the Introduction) suggests that Bohmian mechanics
might not be just one of interpretations, 
but a part of the formalism 
without which the covariant quantum theory cannot be formulated consistently.     
Note also that the adoption of Bohmian mechanics automatically removes 
uncertainties about the interpretational issues  
of quantum theory. 

We also note that, in general, $R^{\mu}$ does not need to be timelike, 
so the preferred foliation does not need to be a foliation into 
spacelike hypersurfaces. This implies that our covariant canonical theory 
is able to deal even with spacetimes that are not globally hyperbolic.

The existence of a dynamically generated preferred foliation 
may also play an important role in semiclassical gravity, especially 
for the problem of definition of particles. This is because 
the definition of particles in the conventional semiclassical gravity 
depends on the choice of time \cite{bd}.

A problem we have not discussed in this paper is how to
describe fermionic fields within the
covariant Bohmian framework. In fact, a ``standard" Bohmian interpretation
of fermionic fields does not yet exist even within the conventional
canonical quantization. However, the work on this issue
is in progress.

To conclude, we believe that the quantization based on De Donder-Weyl 
covariant canonical formalism is an interesting idea 
worthwhile of further investigation. The main advantage is 
the manifest covariance with space and time treated on an equal 
footing. Another advantage is the lack of the interpretational
ambiguities because the Bohmian interpretation emerges
automatically. The main 
problem seems to be the locality/nonlocality issue. 
The approach of \cite{kan} appears to be too local, whereas the approach 
of the present paper allows nonlocalities
that are not allowed by the conventional       
noncovariant Schr\"odinger equation.
This may mean that the whole idea of quantization based on the 
De Donder-Weyl formalism is
wrong, or that it has to be reformulated, or that the nonlocalities
absent in the conventional noncovariant quantization correspond to 
new genuine physical effects.   

%\section*{Acknowledgments}
\vspace{0.6cm}
\noindent
Acknowledgements.
This work was supported by the Ministry of Science and Technology of the
Republic of Croatia under Contract No.~0098002.


\begin{thebibliography}{99}

\bibitem{gsw}
M.B.~Green, J.H.~Schwarz, E.~Witten, Superstring Theory,
Cambridge University Press, Cambridge (1987)
\bibitem{rovlect}
M.~Gaul, C.~Rovelli, Lect.~Notes Phys.~{\bf 541} (2000) 277
\bibitem{thiem}
T.~Thiemann, Lect.~Notes Phys.~{\bf 631} (2003) 41
\bibitem{rovbook}
C.~Rovelli, Quantum Gravity, Cambridge University Press, Cambridge,   
(2004); 
%\begin{verbatim}
http://www.cpt.univ-mrs.fr/$\,\tilde{}\,$rovelli
%\end{verbatim}
\bibitem{banks}
T.~Banks, W.~Fischler, S.H.~Shenker, L.~Suskind,
Phys.~Rev.~D {\bf 55} (1997) 5112
\bibitem{crnwitt}
\v{C}.~Crnkovi\'c, E.~Witten, in: S.W.~Hawking, W.~Israel (Eds.), 
Three Hundred Years of Gravitation, 
Cambridge University Press, Cambridge (1987)
\bibitem{crn}
\v{C}.~Crnkovi\'c, Class.~Quant.~Grav.~{\bf 5} (1988) 1557
\bibitem{kas}
H.A.~Kastrup, Phys.~Rep.~{\bf 101} (1983) 1
\bibitem{got}
M.J.~Gotay, J.~Isenberg, J.E.~Marsden, physics/9801019
\bibitem{rov}
C.~Rovelli, gr-qc/0207043
\bibitem{kan}
I.V.~Kanatchikov, Phys.~Lett.~A {\bf 283} (2001) 25
\bibitem{kan2} 
I.V.~Kanatchikov, Int.~J.~Theor.~Phys.~{\bf 40} (2001) 1121
\bibitem{bohm2}
D.~Bohm, Phys.~Rev.~{\bf 85} (1952) 180
\bibitem{bohmrep}
D.~Bohm, B.J.~Hiley, P.N.~Kaloyerou,
Phys.~Rep.~{\bf 144} (1987) 349
\bibitem{holrep}
P.R.~Holland, Phys.~Rep.~{\bf 224} (1993) 95
\bibitem{holbook}
P.R.~Holland, The Quantum Theory of Motion,
Cambridge University Press, Cambridge (1993)
\bibitem{ajp}
D.F.~Styer et al., Am.~J.~Phys.~{\bf 70} (2002) 288
%\bibitem{nikqg}
%H.~Nikoli\'c, gr-qc/0312063.
\bibitem{nikol1}
H.~Nikoli\'c, quant-ph/0307179
\bibitem{nikol2}
H.~Nikoli\'c, quant-ph/0406173, to appear in Found.~Phys.~Lett.
\bibitem{bar}
J.A.~deBarros, N.~Pinto-Neto, M.A.~Sagioro-Leal, 
Phys.~Lett.~A {\bf 241} (1998) 229
\bibitem{col}
R.~Colistete Jr., J.C.~Fabris, N.~Pinto-Neto, 
Phys.~Rev.~D {\bf 57} (1998) 4707
\bibitem{pn}
N.~Pinto-Neto, R.~Colistete Jr., Phys.~Lett.~A {\bf 290} (2001) 219
\bibitem{mar}
J.~Marto, P.V.~Moniz, Phys.~Rev.~D {\bf 65} (2001) 023516
\bibitem{pn2}
N.~Pinto-Neto, E.S.~Santini, Phys.~Lett.~A {\bf 315} (2003) 36
\bibitem{barb}
G.D.~Barbosa, N.~Pinto-Neto, Phys.~Rev.~D {\bf 69} (2004) 065014
\bibitem{rieth}
J.~von Rieth, J.~Math.~Phys.~{\bf 25} (1984) 1102
\bibitem{val} 
A.~Valentini, Phys.~Lett.~A {\bf 156} (1991) 5
%\bibitem{wein}
%S.~Weinberg, {\it Gravitation and Cosmology} 
%(John Wiley \& Sons, New York, 1972).
\bibitem{hor}
P.~Ho\v rava, Class.~Quant.~Grav.~{\bf 8} (1991) 2069
\bibitem{kan3}
I.V.~Kanatchikov, 
Rept.~Math.~Phys.~{\bf 53} (2004) 181
%hep-th/0301001
\bibitem{kuc} 
K.~Kucha\v r, in: W.~Israel (Ed.),
Relativity, Astrophysics, and Cosmology,
D.~Reidel Publishing Company,
Dordrecht/Boston (1973)
\bibitem{padm}
T.~Padmanabhan, Int.~J.~Mod.~Phys.~A {\bf 4} (1989) 4735
\bibitem{kuc2}
K.~Kucha\v r, in: Proceedings of the 4th Canadian Conference
on General Relativity and Relativistic Astrophysics,
World Scientific, Singapore (1992)
\bibitem{ish} 
C.J.~Isham, gr-qc/9210011
\bibitem{grant}
J.D.E.~Grant, I.G.~Moss, Phys.~Rev.~D {\bf 56} (1997) 6284
\bibitem{padmrep}
T.~Padmanabhan, 
Phys.~Rep.~{\bf 406} (2005) 49
%gr-qc/0311036
\bibitem{shoj}
A.~Shojai, F.~Shojai, Class.~Quant.~Grav.~{\bf 21} (2004) 1
\bibitem{espos}
G.~Esposito, G.~Gionti, C.~Stornaiolo, Nuovo Cim.~{\bf B110} (1995) 1137
\bibitem{hort}
G.~Horton, C.~Dewdney, 
J.~Phys.~A {\bf 37} (2004) 11935
%quant-ph/0407089
\bibitem{bd}
N.D.~Birrell, P.C.W.~Davies, Quantum Fields in
Curved Space, Cambridge Press, NY (1982)


\end{thebibliography}
\end{document}